\documentstyle[aps]{revtex}

\begin{document}
\draft
\title{Van Hove Singularity and D-Wave Pairing in Disordered
                                             Superconductors.}

\author{G. Litak$^{\dag}$, A.M. Martin$^{\S}$, B.L. Gy\"{o}rffy$^{\S}$,
J.F. Annett$^{\S}$ and K.I. Wysoki\'nski$^{\ddag}$.}

\address{$^{\dag}$Department of Mechanics, Technical University of
Lublin, Poland. 
\\
$^{\S}$H.H. Wills Physics Laboratory, University of
Bristol, UK.
\\
$^{\ddag}$Institute of Physics,Maria Curie-Sklodowska University, Lublin,
Poland.}

\date{\today}
\maketitle

\begin{abstract}

We apply the coherent potential approximation (CPA) to 
a simple model for disordered superconductors with $d$-wave pairing. 
We demonstrate that whilst the effectiveness of an electronic Van Hove 
singularity to enhance the transition temperature T$_c$ is reduced by 
disorder it is not eliminated. In fact we give a qualitative account of  
changes in the T$_c$ vs. doping curve with increasing disorder and 
compare  our results with experiments on the 
Y$_{0.8}$Ca$_{0.2}$Ba$_2$(Cu$_{1-c}$Zn$_c$)$_{3}$O$_{7-\delta}$ alloys.
\end{abstract}


According to a widely shared point of view the high transition temperature, 
T$_c$, of the superconducting cuprates is  the consequence of a Van 
Hove singularity, in their normal state electronic structure, enhancing an
otherwise unexceptional, that is to say weak, effective electron-electron 
attraction\cite{Mar97}. Clearly, if correct, this scenario would imply
that relatively modest tinkering with such conventional mechanisms of
pairing as electron-phonon interaction or spin-fluctuation could suffice
to solve the central problem of high temperature superconductivity. This, 
conservative, view\cite{Fri89} is supported by two main observations. 
Firstly, most parameter free, first-principles, calculations find a 
Van Hove singularity near the Fermi energy $\epsilon_F$ in these
materials\cite{andersen,novikov}. Secondly, the experimentally observed 
rise and fall of T$_c$ with doping can be interpreted, very naturally, as 
due to $\epsilon_F$ passing a Van Hove singularity. Of course, there are 
many objections to this, so called, Van Hove 
Scenario \cite{Mar97,Rad93,Pic97}. However, in the light of the emerging 
consensus that the Cooper pairs in these superconductors have $d$--like 
internal symmetry \cite{Ann96} two of  these objections are  particularly 
direct and damaging. The first one  is that in the case of $d$--wave pairing 
the Van Hove mechanism is not effective. As it happens, this has been 
adequately  answered by Newns et al. \cite{New95}. The other one concerns 
the particular sensitivity of $d$--wave pairing to disorder and will be
addressed in the present letter.

In the case of $s$--wave superconductors the influence of scattering by 
defects, which do not break time reversal invariance, on T$_c$ is governed 
by the Anderson's Theorem \cite{And59} and hence the disorder averaged 
density of states, $\overline{n}(\epsilon)$. Thus, the effect of disorder 
on the Van Hove Scenario is merely a question of smearing the structure 
in $n(\epsilon)$ corresponding to the singularity. By contrast for 
$d$--wave superconductors there is no Anderson's Theorem and hence the role 
played by disorder is dramatically different. Not surprisingly, even 
without the complications introduced by  Van Hove singularities the 
problem  has been studied only recently and there are still many open 
questions \cite{Gor83,Lee93,Ner94,Abr96,Sun96,Pok96}.
In what follows, we generalize the Coherent Potential Approximation (CPA) 
for disordered superconductors  \cite{Lit92} to the case of 
$d$--wave pairing. In general, the CPA is a significant improvement on  
beyond the self-consistent Born approximation (SBA) \cite{Abr59}, 
currently in use for disordered $d$--wave superconductors.
Elsewhere \cite{us} we shall discuss in detail the formalism,
differences between the CPA and the  SBA, and various implications for
experiments on high temperature superconductivity. 
However, in this letter we will focus our attention  on  
investigating, for the first time, the viability of the Van Hove Scenario in 
$d$--wave superconductors in the presence of disorder. 

Our arguments will be based on  extended, negative $U_{i,j}$ Hubbard
model on a 
lattice, whose sites are labelled by $i$  and $j$, with random site energies 
$\varepsilon_i$.
Moreover, we shall work in the Hartree--Fock--Gorkov approximation  which 
implies that the usual one particle Green function matrix, in Nambu space, 
at the Matsubara 'frequency' $\epsilon_n= \frac{\pi}{\beta} (2n+1)$  
satisfies:

\begin{equation}
\sum_l \left( \begin{array}{c} (i \epsilon_n -\varepsilon_i + \mu) 
\delta_{il}+t_{il}~~~~~\Delta_{ij} \\ 
 \Delta_{ij}^*~~~~~ (i \epsilon_n +\varepsilon_i - \mu) \delta_{il}
-t_{il} \end{array} \right)
 \left( \begin{array}{c} G_{11}(l,j;\epsilon_n)~~ G_{12}(l,j;\epsilon_n) \\
G_{21}(l,j;\epsilon_n)~~ G_{22}(l,j;\epsilon_n) \end{array} \right) = 
\delta_{ij} \left(
\begin{array}{c} 1~~ 0
\\ 0~~ 1  \end{array} \right)~~,
\end{equation}

\noindent where  the hopping integrals $t_{ij}$ and the pairing potentials 
$\Delta_{ij}$ will be taken to 
be non zero only when the sites $i$ and $j$ are nearest neighbours, $\mu$ 
is the 
chemical potential, and we shall often refer to the Greens function matrix  
as   $\underline{\underline{G}}(i,j;\epsilon_n)$. The above equations are 
completed by the self-consistency condition:

\begin{equation}
\Delta_{ij}= U_{ij} \frac{1}{\beta} \sum_n {\rm e}^{i \epsilon_n \eta} 
G_{12}(i,j;\epsilon_n)~~,
\end{equation}

\noindent where $U_{ij}$ is an attractive interaction energy between two 
electrons on nearest 
neighbour sites  and    $\eta$    is a positive infinitesimal. To simplify 
matters we 
have assumed that the Hartree term can be  absorbed into the hopping 
integral  $t_{ij}$  and dropped it from equation (1). As usual equations
(1) and (2) are to be solved 
subject to the requirement  on the chemical potential that

\begin{equation}
n= \frac{2}{\beta} \sum_n {\rm e}^{i \epsilon_n \eta} 
G_{11}(i,i;\epsilon_n)~~, 
\end{equation}

\noindent where $n$ is the number of electrons per unit cell. 

Whilst, later, we shall refer to experiments on the substitutional alloys 
Y$_{0.8}$Ca$_{0.2}$Ba$_2$(Cu$_{1-c}$Zn$_c$)$_{3}$O$_{7-\delta}$ 
\cite{Ber96} we do not wish to  be very specific about the physical 
nature of the point defects represented by the site energies $\varepsilon_i$. 
Indeed, we are content to provide a reliable analysis of the simplest 
possible non trivial model. Thus we take them to be independent random 
variables defined to have values  $\frac{1}{2} \delta$ and  
$-\frac{1}{2} \delta$ with equal probability, $\frac{1}{2}$, on every site. 
As might be expected we shall be interested in the average of 
$\underline{\underline{G}}(i,j;\epsilon_n)$ over the above ensemble. To 
calculate $\underline{\underline{\overline G}}(i,j;\epsilon_n)$ we shall 
make use of the Coherent Potential Approximation  which is well known to 
be the mean field theory of disorder for problems similar to the one at 
hand \cite{CPA}.

To define a tractable problem of interest we shall  assume that the sites 
form a square lattice. Then for $\varepsilon_i=0$ for all $i$, in the normal
state, where $\Delta_{ij}=0$, the spectrum features 4, well known, Van Hove 
singularities at $\epsilon=0$, resulting in logarithmic divergence: 
$n(\epsilon) \approx -ln(\epsilon)$. Below a certain temperature T$_c$
there is a solution with  $\Delta_{ij} \ne 0$ and $d$--wave symmetry. As 
was pointed out by Newns et al. \cite{New95} in this model when $n$ 
changes the Van Hove Scenario  obtains as  readily as in models where the 
symmetry of superconducting state is $s$  type. That is to  say T$_c$  
rises to a  maximum at $n=1$ and then falls as $n$ increases from $n=0$ to 
$n=2$ for a fixed interaction strength  $U_{ij}$. As mentioned above, 
the technical question we shall answer in this letter is {\it  what happens 
to this behaviour when the site energies are not zero but randomly}
 $\frac{1}{2} \delta$ {\it or} $-\frac{1}{2} \delta$?

Clearly, the relevant, generic, consequence of disorder is the smearing of 
structure in the density of states $\overline n (\epsilon )$. To gauge the 
extent of this in our model we calculated $\overline n (\epsilon )$  
using the standard CPA procedure \cite{CPA}. The results, illustrating 
the smearing of the Van Hove singularities at $\epsilon=0$, are shown in 
Fig. 1(a) for various values of the scattering strength $\delta$. In 
Fig.1(b) we also show the corresponding self energy $\Sigma ( \epsilon)$ 
which in CPA depends only on $\epsilon$ but not on the wave vector 
$\vec k$. For our simple model $\Sigma(\epsilon_n)$, at the Matsubara 
frequency $\epsilon_n$, satisfies the following CPA equation \cite{CPA}

\begin{equation}
\Sigma(\epsilon_n) = (\frac{1}{2} \delta - \Sigma(\epsilon_n) ) 
\overline G (i,i; \epsilon_n)
(\frac{1}{2} \delta + \Sigma(\epsilon_n) )~~.
\end{equation}

\noindent It will be useful to note that in the weak scattering limit 
equation (4) reduces to the Self-consistent Born Approximation (SBA)

\begin{equation}
\Sigma^{SBA} (\epsilon_n) = \frac{\delta^2}{4} \overline G (i,i; \epsilon_n)
\end{equation}

\noindent and hence on the real energy axis  
${\rm Im} \Sigma(\epsilon) \sim n(\epsilon)$. Interestingly, in the 
non-self-consistent Born approximation ${\rm Im} \Sigma(\epsilon)$ is 
logarithmically divergent at the van Hove singularity. In the disordered 
case one would expect such a singularity to be smeared out into a peak. 
Indeed the $\delta=0.6$ curve in Fig. 1(b) is fully consistent with this 
expectation. The other curves show the significant deviation, the split 
Van Hove singularity for instance, between CPA and SBA.

Let us now turn to the case where both superconductivity and disorder are 
present\cite{Lit92}. Although the full CPA program can be implemented for the 
problem defined by equations (1-3) and the specification of the site 
energy ensemble, it is convenient to make the approximation, valid when
the coherence length $\xi_0$ is much larger then the lattice spacing, that 
the pairing potential $\Delta_{ij}$ does not fluctuate very much and replace 
it in equation (1) by its average value $\overline \Delta_{ij}$ \cite{Gy10}. 
For conventional $s$--wave pairing this leads to the Anderson's Theorem 
\cite{And59} which means that the only effect of disorder is to replace the 
density of states in the gap equation by its ensemble average  
$\overline n(\epsilon)$ \cite{Gy10}. Thus, in the Van Hove Scenario it is
clear that the smearing  of  the  Van Hove singularity  in Fig. 1(a)  
implies a weakening of the Van Hove enhancement of 
${\rm T}_c \sim exp[-1/(U \overline n(\epsilon_F))]$. The T$_c$ at optimal 
doping will thus be significantly reduced by disorder, even for $s$--wave 
pairing.

As mentioned earlier the theory of disorder in $d$--wave superconductors 
turns out to be very different \cite{Gor83,Lee93,Ner94,Abr96,Sun96,Pok96}.
Nevertheless, the above simplification remains both valid and useful. In 
short we shall  deploy the CPA  to find    
$\underline{\underline{ \overline G}}(i,j;\epsilon_n)$ for an averaged
pairing potential $\overline \Delta_{ij}$ (old) and recalculate 
$\overline \Delta_{ij}$ (new) using equation (2) with 
$G_{12}(i,j;\epsilon_n)$ replaced by $\overline G_{12}(i,j;\epsilon_n)$ 
repeating  the process until convergence. In this way self-consistency on 
the average is ensured. 

To  derive the  basic CPA equations for disordered $d$--wave superconductors 
let us define the coherent Greens function  
$\underline{\underline{ G}}^c(i,j;\epsilon_n)
\equiv \underline{\underline{ \overline G}}(i,j;\epsilon_n)$ by the equation:

\begin{equation}
\sum_l \left( \begin{array}{c} (i \epsilon_n + \mu-\Sigma_{11}(\epsilon_n)) 
\delta_{il}+t_{il}~~~~~
\overline \Delta_{il} \\
\overline \Delta_{il}^*~~~~~ (i \epsilon_n - \mu -\Sigma_{22}(\epsilon_n)) 
\delta_{il}
-t_{il} \end{array} \right) \underline{\underline{ G}}^c(l,j;\epsilon_n)
 = \delta_{ij} \left(
\begin{array}{c} 1~~ 0
\\ 0~~ 1  \end{array} \right)~~,
\end{equation}

\noindent where we did not  introduce any off diagonal self-energies such as 
$\Sigma_{12}(\epsilon_n)$  and $\Sigma_{21}(\epsilon_n)$
because for  the single site perturbations of our model they would turn out 
to be zero. The next step is to consider the scattering of the 
quasi-particles propagating according to $G^c(i,j;\epsilon_n)$ by the 
defects described by the potentials:

\begin{equation}
\underline{\underline{V}}^{\pm} = \left( \begin{array}{c} \pm \frac{1}{2} 
\delta~~~  0 \\
0~~~ \mp \frac{1}{2} \delta \end{array} \right) - \left( \begin{array}{c}
\Sigma_{11}(\epsilon_n)~~~
0 \\
0~~~ \Sigma_{22}(\epsilon_n) \end{array} \right)~~.
\end{equation}

In a straightforward application of the CPA principles
$\underline{\underline{\Sigma}}(\epsilon_n)$ and therefore 
$\underline{\underline{G}}^c(i,j;\epsilon_n)$ is determined by the condition 
that these defects do not scatter on the average. After some algebra this 
leads to the condition, similar to the one in equation (4), that:

\begin{equation}
\Sigma_{11}(\epsilon_n) = (\frac{1}{2} \delta - \Sigma_{11}(\epsilon_n) ) G^c_{11} (i,i; \epsilon_n)
(\frac{1}{2} \delta + \Sigma_{11}(\epsilon_n) )~~.
\end{equation}

We solved this equation numerically by iteration  using the fact  that  
$G_{11}^c(i,j;\epsilon_n)=G_{11}^0(i,j;\tilde \epsilon_n,\tilde \mu)$ 
where  $\underline{\underline{G}}^0$ is the solution of equation (1)
with $\varepsilon_i=0$ for every $i$, $\Delta_{ij}= \overline \Delta_{ij}$ 
and the renormalized frequencies   $\tilde \epsilon_n$ and chemical
potential $\tilde \mu$ are given by

\begin{equation}
\tilde \epsilon_n = \epsilon_n - {\rm Im} 
\Sigma_{11}(\epsilon_n)~~,~~~~~~\tilde \mu= \mu + {\rm Re}
\Sigma_{11}(\epsilon_n)~~.  
\end{equation}

\noindent Note that equation (9) together with  equation (8) constitutes a
closed loop of relations which determine $\tilde \epsilon_n$  or 
alternatively $\Sigma_{11}(\epsilon_n)$. Clearly, to solve this we 
need the Greens function matrix $\underline{\underline{G}}^0(i,i;\epsilon_n)$.
This was obtained numerically by a very efficient recursion method 
\cite{Mar98}.

To place above formulae in the context of previous work we note that in the  
weak scattering limit they reduce to those investigated in references 
\cite{Gor83,Abr96,Sun96,Pok96} where $\Sigma_{11}(\epsilon_n)$ was obtained 
by the SBA in equation (5). By contrast  the CPA resums diagrams to all 
orders including some which have been investigated by Nersesyan et al. 
\cite{Ner94}. However, unlike them we find, analytically,  that, for small 
$\delta$, near the Fermi energy 
$\epsilon \approx 0$, 
$n(\epsilon) \approx \frac{16\Delta}{\pi\delta^2}
{\rm e}^{\frac{-8 \pi \Delta t}{\delta^2}} \ne 0$, as in the earlier 
work of Gorkov and Kalugin \cite{Gor83}.

In Fig. 2 we display our results for the evolution of the density of states 
$n(\epsilon)$ and the self-energy  $\Sigma(\epsilon)$ with disorder, 
as described by $\delta$, at $T=0$. As expected the overall effect of 
disorder is to fill in the v--shaped dip in the density of states near 
$\epsilon_F$. However, this process is rather intricate due to the competition 
between contributions in $k$--space from near the point where 
$\Delta(\vec k)=0$ and the Van Hove singularities at the saddle points. As 
a result, for small $\delta$, the strong scattering at the Van Hove 
singularity, evident in Fig. 1(b), is  suppressed in the superconducting
state by the dip,  but eventually, for large $\delta$, disorder wins.
Namely  $\Sigma_{11}(0)$ recovers and the gap is suppressed. Interestingly, 
the size of the `gap' as measured by the distance between the two peaks in 
Fig. 2(a) remains roughly constant as the low energy dip is filled in. 
We have also performed finite temperature calculations relevant to tunnelling 
experiments \cite{Tun}, but we shall present these in a separate publication 
\cite{us}.

Finally we investigate the dependence of T$_c$ on the band filling $n$.
Our results for T$_c$ vs. $n$ are displayed in Fig. 3(a). For orientation we
also calculated  T$_c$ as a function of $n$ for an on site  only, negative 
$U$   Hubbard model  with  $U_{ij}=-|U|\delta_{ij}$ and the same $|U|$ and  
hopping integrals $t_{ij}$ on the same lattice. The corresponding results are 
displayed in Fig. 3(b). Surprisingly, in spite of the delicate interaction 
between the Van Hove singularities  and the very different behaviour of the 
quasiparticle spectra at low energies the two curves are largely similar. 
Hence, we may conclude that the Van Hove Scenario survives disorder  almost 
as readily  in  $d$--wave superconductors as in the more conventional 
$s$--wave case. In support of the suggestion that this fact might have 
something to do with  the superconducting cuprates we reproduce, in Fig. 4 
the measurements of T$_c$, as a function of oxygen deficiency for various 
concentrations of Zn, made on 
Y$_{0.8}$Ca$_{0.2}$Ba$_2$(Cu$_{1-y}$Zn$_y$)$_3$O$_{7-\delta}$ samples by 
Bernhard et al. \cite{Ber96}. Clearly if the Zn concentration is regarded 
of a measure of disorder and oxygen deficiency as that of doping these 
curves are very similar to  our theoretical results in Fig 3(a). Thus our 
calculations support the interpretation of the experimental data by Bernhard 
et al. \cite{Ber96} as evidence for $d$--wave pairing.
\bigskip

\noindent {\bf Acknowledgements:}
This work has been partially supported by the British -- Polish
Joint Research Collaboration Programme (WAR/992/076, DZ/2631/JP)
and by the EPSRC under grant number GR/L22454 and KBN grant 2P 03B 031 11.

\begin{figure}
\caption{Density of states $N(E)$ (a) and self energies $\Sigma (E)$ (b)
for a normal state with various disorder strengths $\delta$.}
\end{figure}

\begin{figure}
\caption{Density of states $N(E)$ (a) and self energies $\Sigma (E)$ (b) for a
superconducting state with various disorder strengths $\delta$, 
calculated for $|U|=3.5t$ and $n=1$.}
\end{figure}

\begin{figure}
\caption{Critical temperature T$_c$ vs. band filling $n$ for $d$ (a)
and $s$--wave (b) superconductors and a number of disorder strengths
$\delta$ ($|U|=3.5t$).}
\end{figure}

\begin{figure}
\caption{T$_c$ as a function  of oxygen deficiency $\delta$ for
Y$_{0.8}$Ca$_{0.2}$Ba$_2$(Cu$_{1-y}$Zn$_y$)$_3$O$_{7-\delta}$ with
$y=0$ (full squares), $y=0.02$ (triangles ), $y=0.04$ (circles) and 
$y=0.06$ (stars) [19].}
\end{figure}

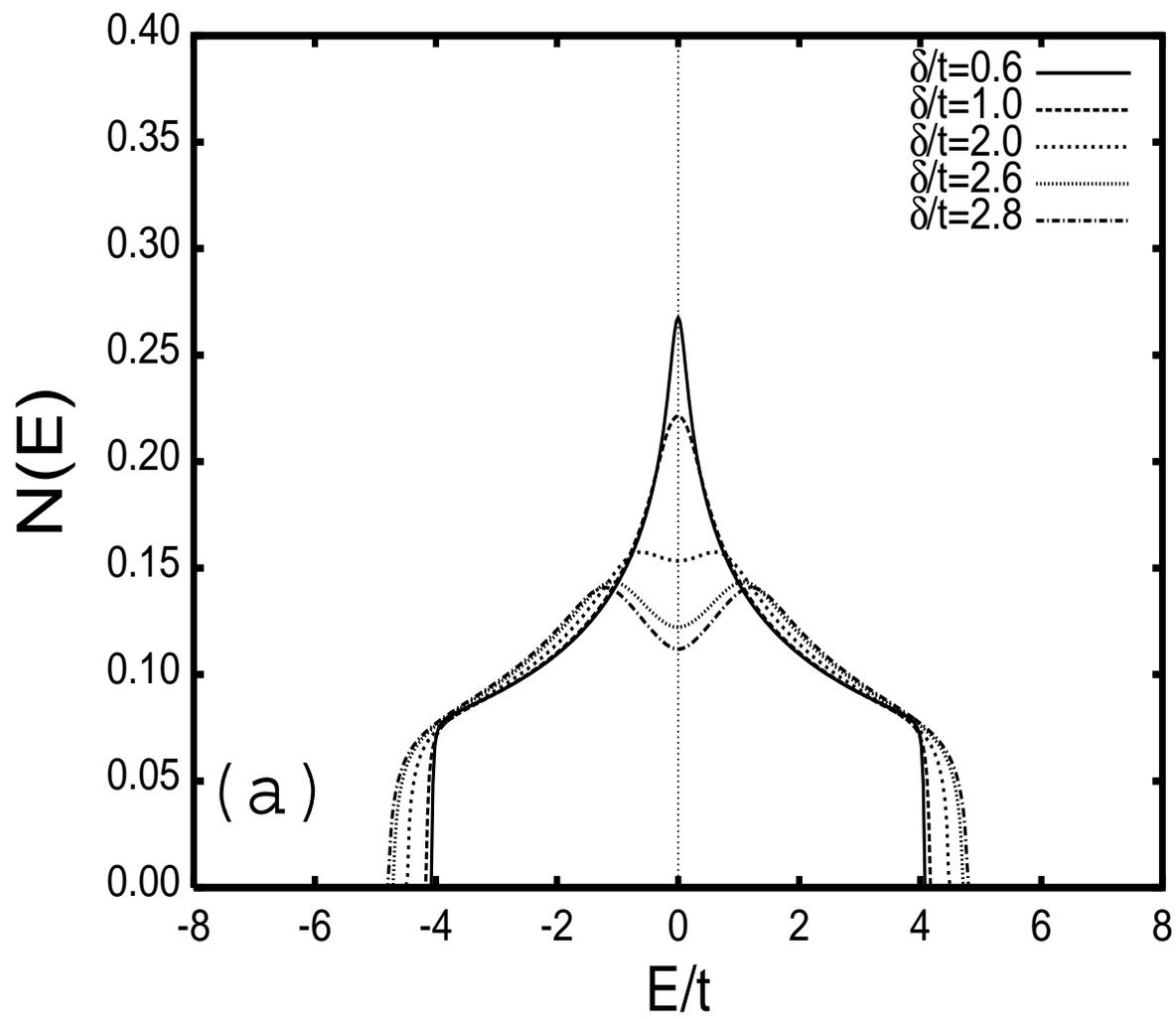
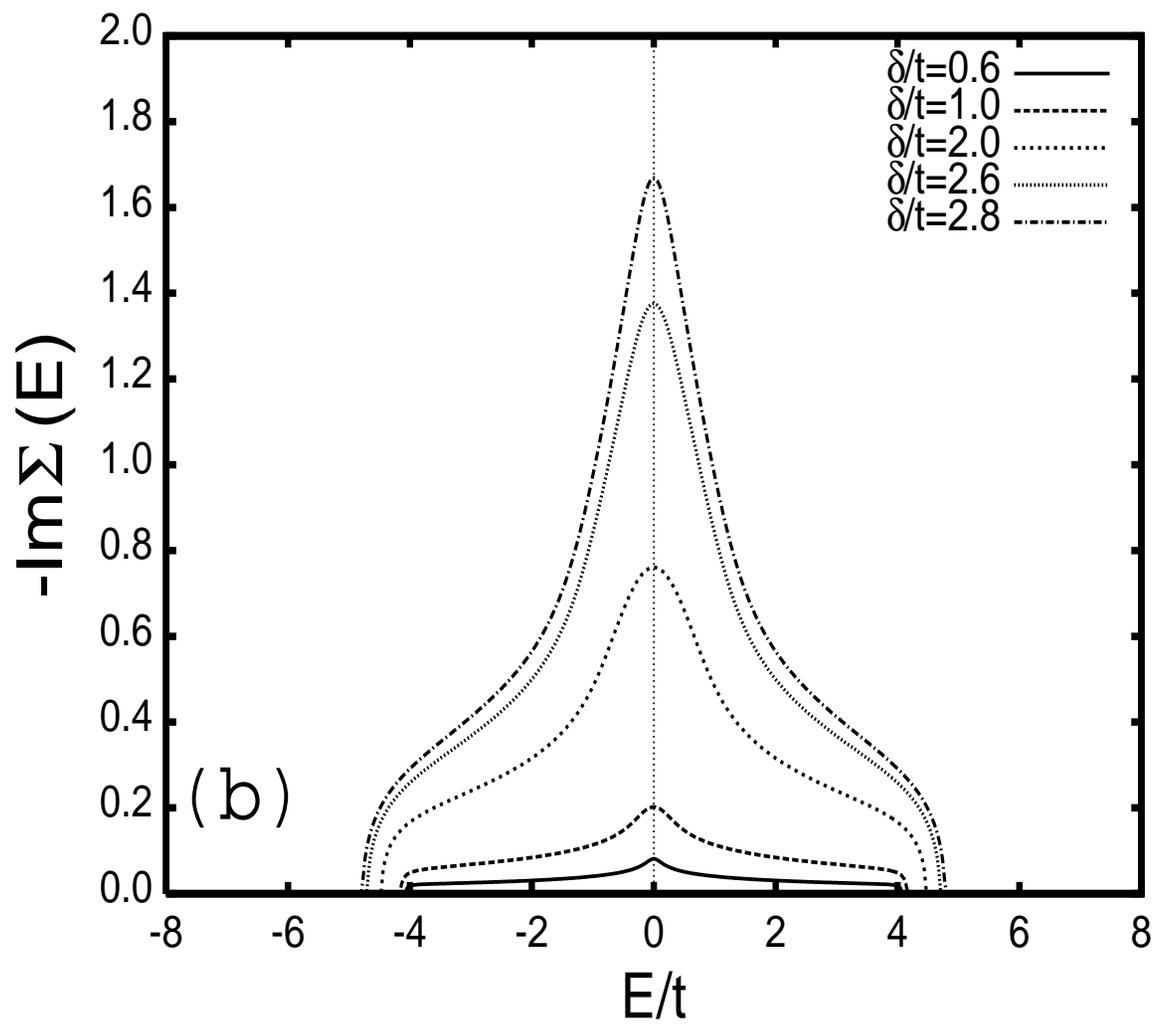
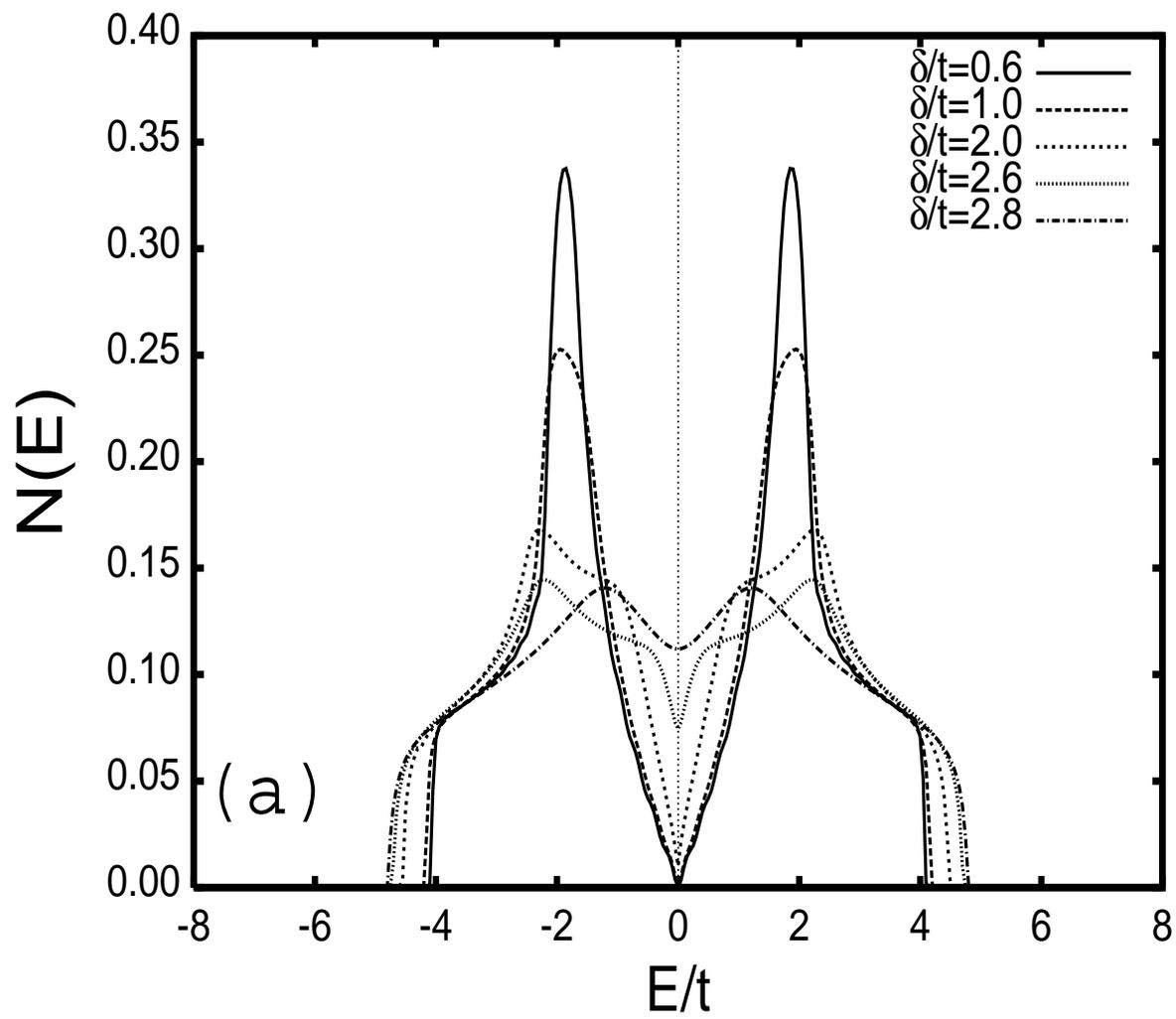
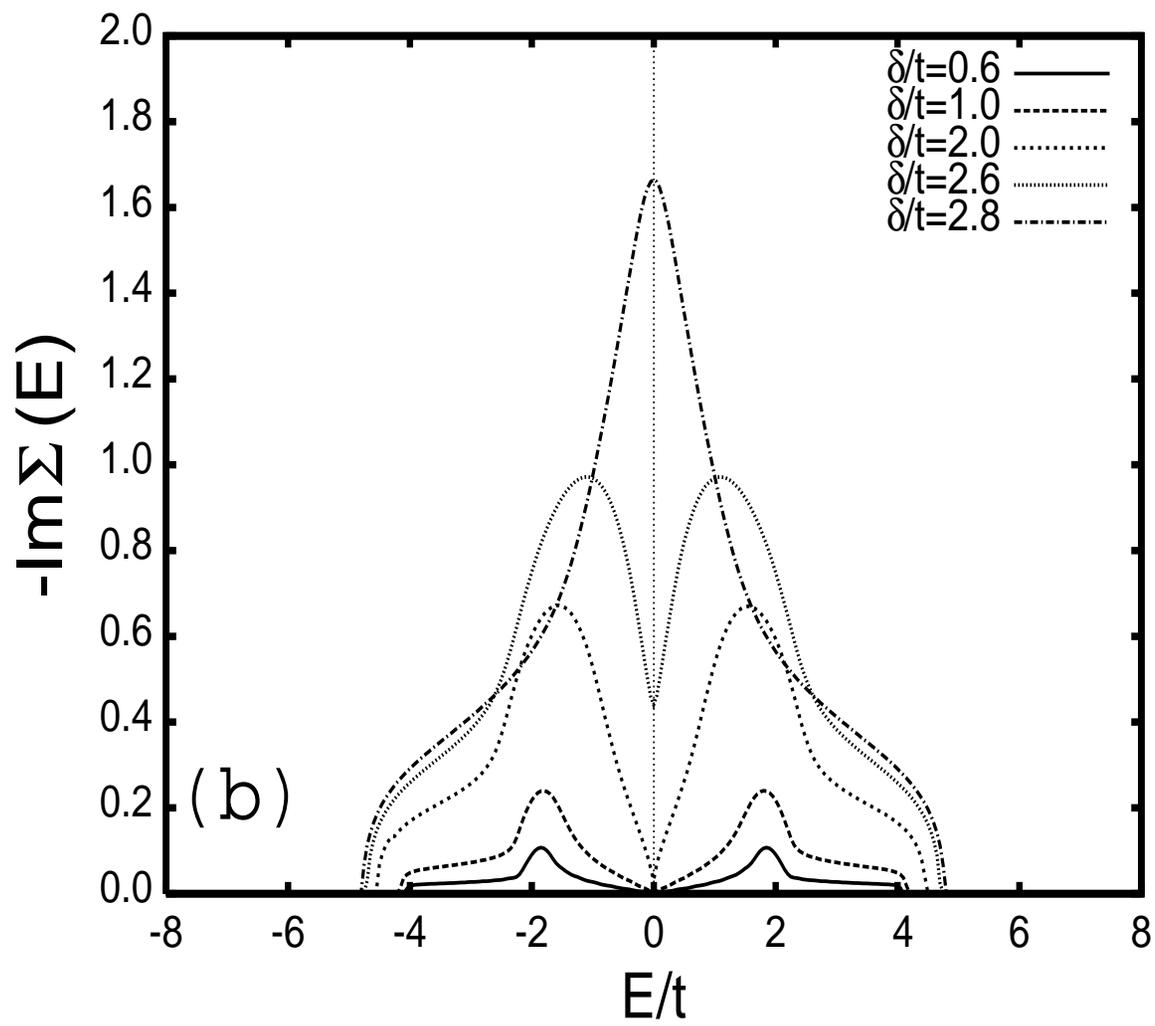
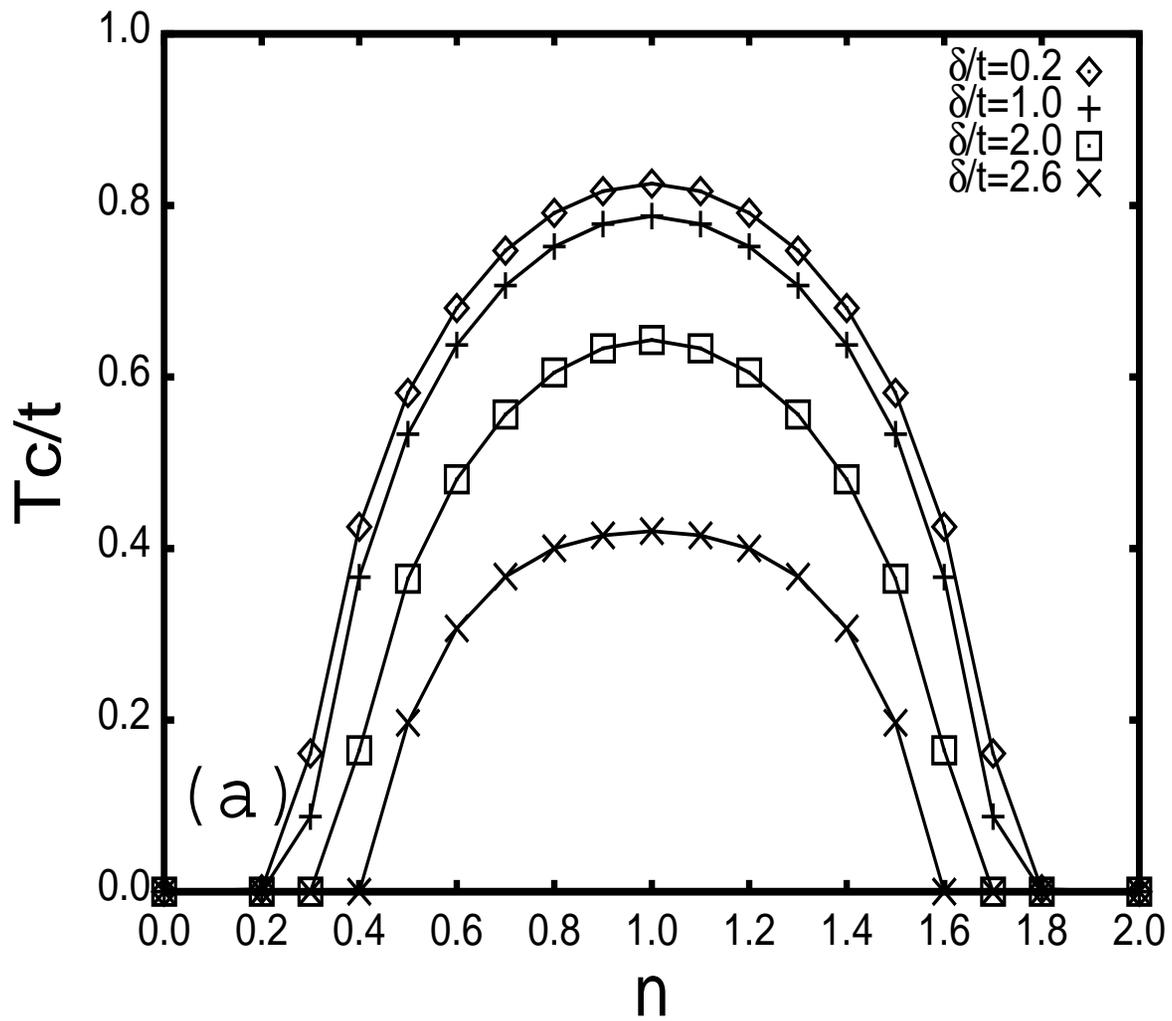
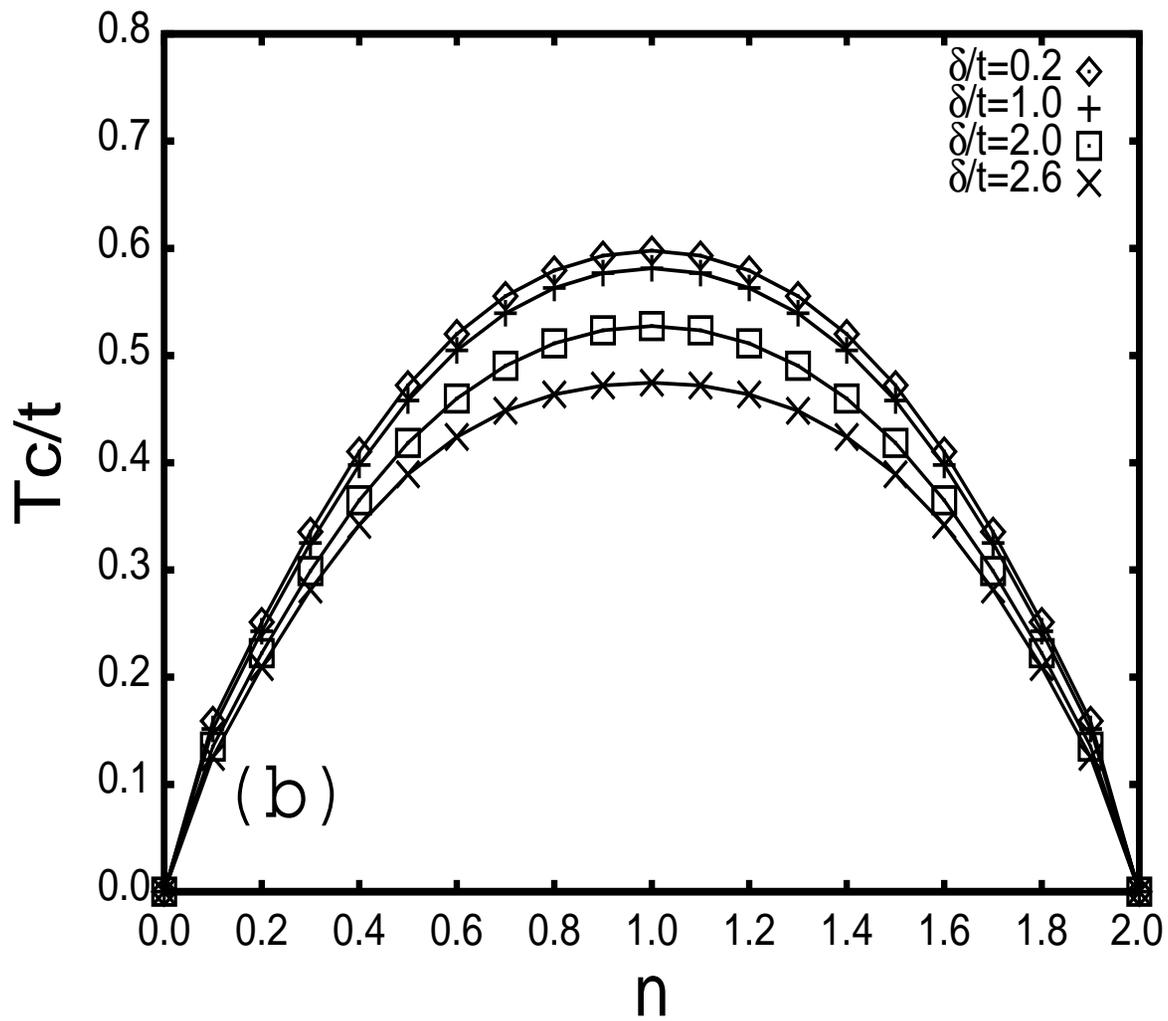
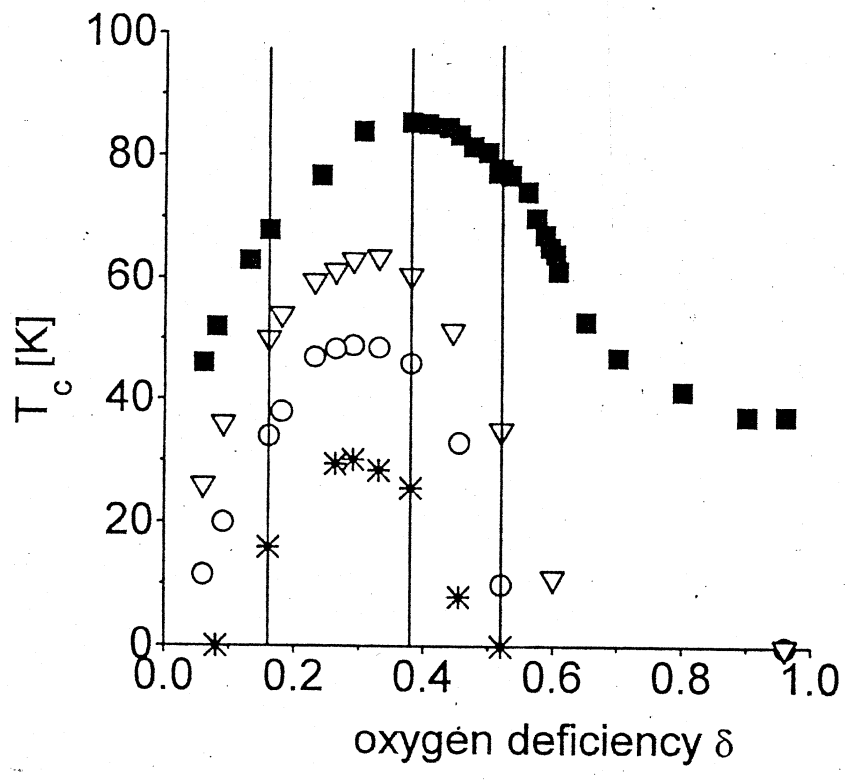
\end{document}